\documentclass[showkeys,floatfix, noshowpacs, preprintnumbers, twocolumn, amsmath, amssymb, aps, superscriptaddress, pra, 10pt]{revtex4-1}
\pdfoutput=1

\usepackage{amsmath,amsthm,amssymb,amsfonts}

\usepackage{eqnarray}

\usepackage{graphicx, soul}
\usepackage{multirow}
\usepackage{url}
\usepackage[usenames,dvipsnames]{xcolor}
\usepackage[breaklinks=true]{hyperref}
\usepackage{bbold}
\usepackage{bm}

\begin{document}

\title{Experimental multiphase estimation on a chip}

\author{Emanuele Polino}
\affiliation{Dipartimento di Fisica, Sapienza Universit\`{a} di Roma, Piazzale Aldo Moro 5, I-00185 Roma, Italy}

\author{Martina Riva}
\affiliation{Istituto di Fotonica e Nanotecnologie, Consiglio Nazionale delle Ricerche (IFN-CNR), Piazza Leonardo da Vinci, 32, I-20133 Milano, Italy}
\affiliation{Dipartimento di Fisica, Politecnico di Milano, Piazza Leonardo da Vinci, 32, I-20133 Milano, Italy}

\author{Mauro Valeri}
\affiliation{Dipartimento di Fisica, Sapienza Universit\`{a} di Roma, Piazzale Aldo Moro 5, I-00185 Roma, Italy}

\author{Raffaele Silvestri}
\affiliation{Dipartimento di Fisica, Sapienza Universit\`{a} di Roma, Piazzale Aldo Moro 5, I-00185 Roma, Italy}

\author{Giacomo Corrielli}
\affiliation{Istituto di Fotonica e Nanotecnologie, Consiglio Nazionale delle Ricerche (IFN-CNR), Piazza Leonardo da Vinci, 32, I-20133 Milano, Italy}
\affiliation{Dipartimento di Fisica, Politecnico di Milano, Piazza Leonardo da Vinci, 32, I-20133 Milano, Italy}

\author{Andrea Crespi}
\affiliation{Istituto di Fotonica e Nanotecnologie, Consiglio Nazionale delle Ricerche (IFN-CNR), Piazza Leonardo da Vinci, 32, I-20133 Milano, Italy}
\affiliation{Dipartimento di Fisica, Politecnico di Milano, Piazza Leonardo da Vinci, 32, I-20133 Milano, Italy}

\author{Nicol\`o Spagnolo}
\affiliation{Dipartimento di Fisica, Sapienza Universit\`{a} di Roma, Piazzale Aldo Moro 5, I-00185 Roma, Italy}

\author{Roberto Osellame}
\affiliation{Istituto di Fotonica e Nanotecnologie, Consiglio Nazionale delle Ricerche (IFN-CNR), Piazza Leonardo da Vinci, 32, I-20133 Milano, Italy}
\affiliation{Dipartimento di Fisica, Politecnico di Milano, Piazza Leonardo da Vinci, 32, I-20133 Milano, Italy}

\author{Fabio Sciarrino}
\affiliation{Dipartimento di Fisica, Sapienza Universit\`{a} di Roma, Piazzale Aldo Moro 5, I-00185 Roma, Italy}
\email{fabio.sciarrino@uniroma1.it}

\begin{abstract}
Multiparameter estimation is a general problem that aims at measuring unknown physical quantities, obtaining high precision in the process. In this context, the adoption of quantum resources promises a substantial boost in the achievable performances with respect to the classical case. However, several open problems remain to be addressed in the multiparameter scenario. A crucial requirement is the identification of suitable platforms to develop and experimentally test novel efficient methodologies that can be employed in this general framework. We report the experimental implementation of a reconfigurable integrated multimode interferometer designed for the simultaneous estimation of two optical phases. We verify the high-fidelity operation of the implemented device, and demonstrate quantum-enhanced performances  in two-phase estimation with respect to the best classical case, post-selected to the number of detected coincidences. This device can be employed to test general adaptive multiphase protocols due to its high reconfigurability level, and represents a powerful platform to investigate the multiparameter estimation scenario.
\end{abstract}

\maketitle

\section{Introduction}

Quantum metrology aims at exploiting quantum resources to enhance the sensitivity in the estimation of unknown physical parameters with respect to what can be achieved with classical approaches \cite{helm,Holevo}. This field of research is increasingly active and represents one of the most promising applications of quantum information theory \cite{giova1,giova2,giova3,pezze}. In the single parameter case, the estimation of an unknown physical quantity with classical resources is bounded by the standard quantum limit (SQL), stating that the achievable error on the unknown parameter scales as $\nu^{-1/2}$, being $\nu$ the number of particles. Such limit can be improved by adopting quantum resources, defining the more fundamental Heisenberg limit (HL) scaling as $\nu^{-1}$ \cite{giova1,giova2,giova3}. Recently, the first unconditional violation of the SQL was reported in \cite{slussa}. Given a probe preparation, the optimal limit for single parameter estimation can be always saturated by appropriately choosing the performed measurement \cite{braun}, and thus the HL effectively represents the ultimate achievable precision limit.

\begin{figure*}[ht!]
\centering
\includegraphics[width=0.99\textwidth]{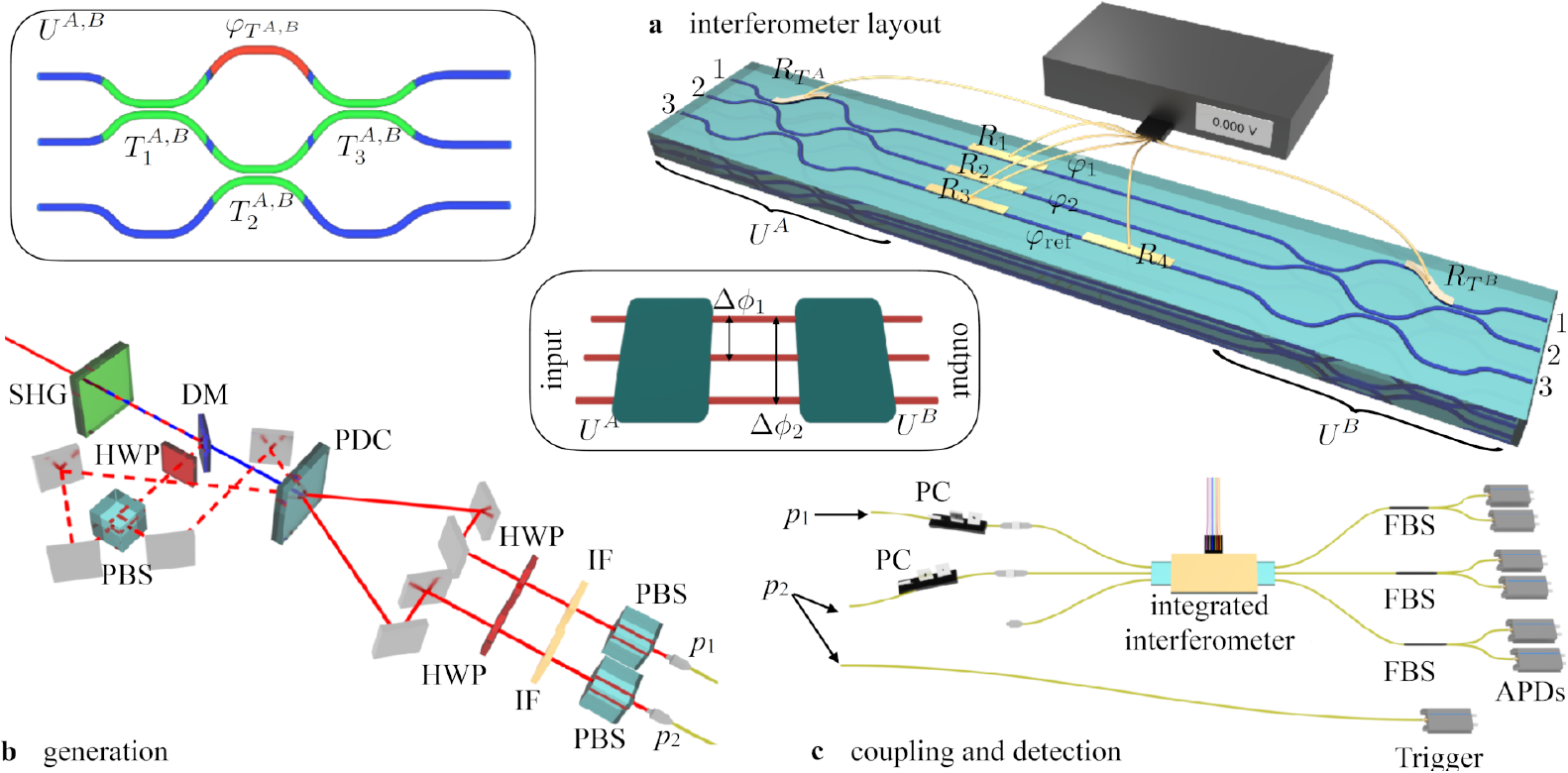}
\caption{Experimental apparatus. {\bf a}, Layout of the integrated reconfigurable device. Three straight waveguide segments are included between two multiport splitters $U_A$ and $U_B$. The dynamical control of the phases is achieved by thermo-optic phase shifters. Central inset: conceptual scheme of the interferometer. Top left inset: layout of the multiport splitters $U^{A,B}$, each composed of three directional couplers ($T_{1,2,3}^{A,B}$, green regions) and a dynamically reconfigurable phase ($\varphi_{T^{A,B}}$, red). By appropriately tuning $\varphi_{T^{A,B}}$ the two multiport splitters can be set to operate as balanced tritters. {\bf b}, Parametric down-conversion source for generation of single-photon and two-photon states. The dotted path is employed to inject classical light into the device for the device alignment. The generated photons ($p_1$ and $p_2$) are coupled in single mode fibers and sent to the integrated device. {\bf c}, Coupling and detection stage. Photons are coupled to the device by an input fiber array (single-mode operation), and collected with a second fiber array (multimode operation). For single-photon inputs, photon ($p_2$) is directly measured to act as a trigger. For two-photon inputs, both photons are injected in the interferometer, and the output state is measured by adding a set of fiber beam-splitters to detect bunching events. (PDC: parametric down-conversion, SHG: second harmonic generation, DM: dichroic mirror, HWP: half wave plate, PBS: polarizing beam-splitter, IF: interference filter, PC: polarization controller, FBS: fiber beam-splitter, APD: avalanche photodiode.)}
\label{fig:device}
\end{figure*}

A natural generalization of quantum metrology aims at extending such results to the simultaneous estimation of more than a single parameter. Indeed, the capability of obtaining quantum-enhanced performances in the multiparameter case is particularly relevant \cite{Datta}, since a large variety of estimation problems involve more than a single physical quantity. Notable examples are phase imaging \cite{Tsang,giov,shin}, measurements on biological systems \cite{Taylor,crespiAPL}, magnetic field imaging \cite{Pham}, gravitational waves parameters estimation \cite{freise,Sch}, sensing technologies \cite{youse,Ivanov}, quantum sensing networks \cite{proc}, quantum process tomography \cite{Zhou,Kahn,Acin,haya} and state estimation \cite{yang}.

Although multiparameter estimation holds a broad range of applications, there are still several open questions with respect to the single parameter case. For instance, while the theoretical framework for the single-parameter scenario is well established \cite{Pari09}, few recipes to saturate the ultimate bounds are known in the multiparameter case \cite{helm,hels,yuen,matsu,fuji,pezze17}. Due to possible non-commutativity of the quantum measurements required to simultaneously optimize the estimation of different parameters \cite{Barn}, it may not be possible to optimally estimate all parameters at the same time. In the case of $d$ compatible parameters, a reduction of resources by a factor $d$ can be obtained with respect to single individual estimations \cite{ragy}. On the one side, simultaneous multiparameter estimation can surpass the individual optimized strategies, but the definition of general quantum bounds still requires additional investigations. On the other side, several physical processes are characterized by dynamics that require intrinsically the simultaneous treatment of all relevant parameters.  

In the last few years, several theoretical investigations on multiparameter estimation have been reported \cite{Datta,Nich,gao,saraf,Humphreys,pezze17,Liu2,Yuan}, while experimental tests are surprisingly few. These include the simultaneous estimation of phase and its diffusion noise \cite{vidri,roccia,alto}, phase and quality of the probe state \cite{roccia18}, the discrimination of an actual signal from parasitic interference \cite{stein}, and quantum-enhanced tomography of an unknown unitary process by multiphoton states \cite{Zhou}.

\begin{figure*}[ht!]
\centering
\includegraphics[width=1\textwidth]{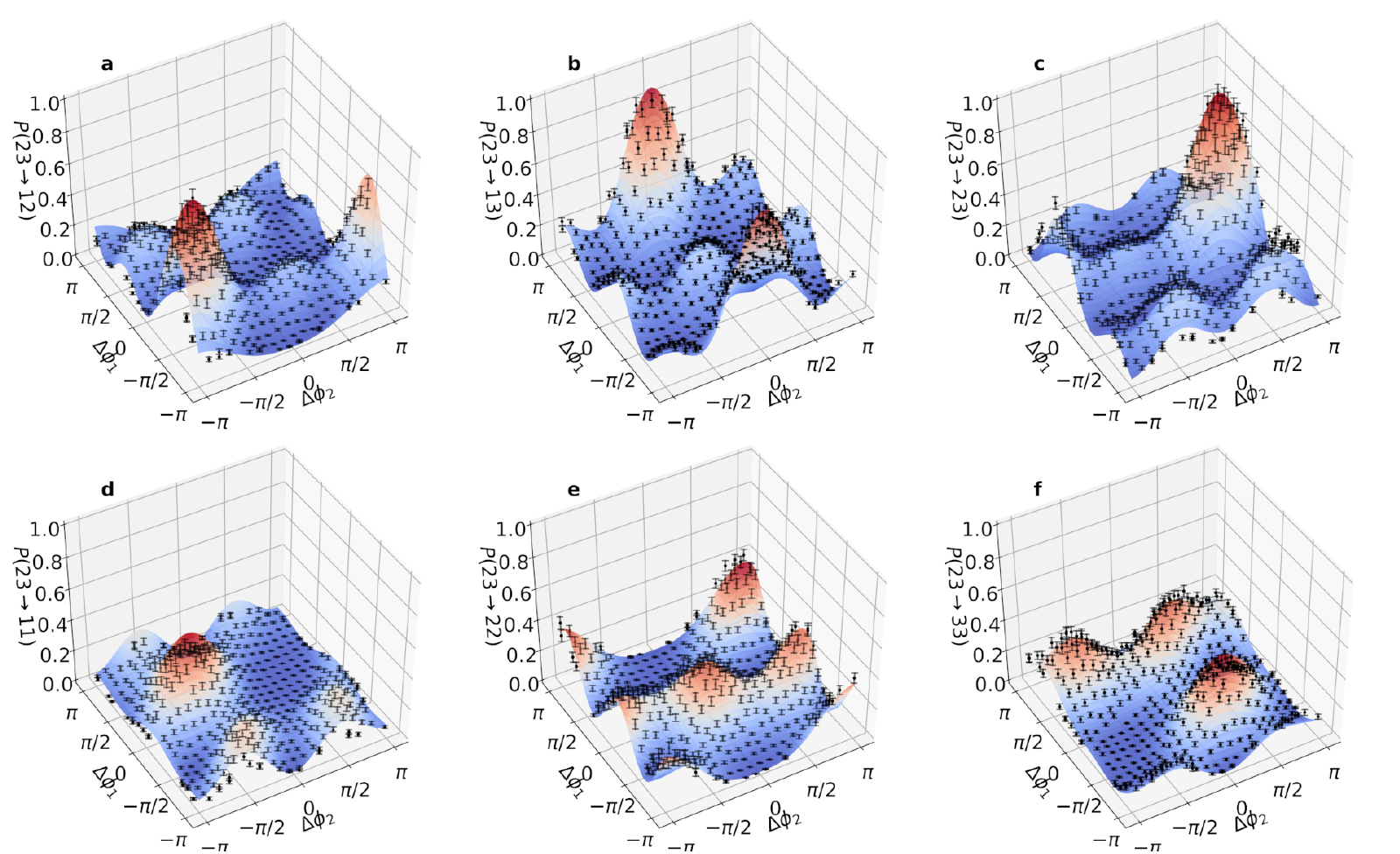}
\caption{
{\bf a-f}, Two-photon probabilities $P(23 \rightarrow ij)$ as a function of the phase differences $\Delta \phi_1$ and $\Delta \phi_2$. The latter are varied changing the dissipated powers on resistors $R_1$ and $R_2$. In all plots, dots are experimental data while surfaces are the theoretical expectations from the circuit characterization process. Error bars are standard deviations due to the Poissonian statistics of the measured single-photon counts and two-photon coincidences. The good agreement between model and experimental data is quantified by the average $R^{2}$ value over all output combinations $\langle R^{2} \rangle = 0.835$. In the model, photon indistinguishability of $V=0.95$ is taken into account.}
\label{3Dplot}
\end{figure*}

It is thus crucial to identify a specific scenario and a corresponding suitable experimental platform to investigate multiparameter estimation tasks. Such scenario is provided by the multiphase problem, where the parameters to be estimated are a set of optical phases. Several theoretical works have been reported in this direction \cite{gaga,pezze17, ciamp,pezze2,proc,Humphreys,Spagnolo13,pezze,macchia,liu, Ge, knott1,Yang2,Zhang1}. Very recent results reported necessary and sufficient conditions to define the optimal projective measurements for pure states \cite{pezze17}, with a subsequent extension to general probe states \cite{Yang2}. Furthermore, generalized matrix bounds and optimal states have been defined in \cite{Gessner}. Nevertheless, no experimental realizations have been reported yet. The most suitable platform to implement multiphase estimation tasks is provided by integrated multiarm interferometers injected by multiphoton states \cite{ciamp}. Such platform presents several advantages in terms of stability, tunability and compactness of the devices \cite{Flam15,Carolan15,Harris17,Chaboyer}.

In this work we report on an integrated three-mode interferometer built through the femtosecond laser writing (FLW) technique \cite{osellameBook,crespiAPL}, to implement quantum multiphase estimation  tasks. Such device is composed by two cascaded tritters \cite{spag13} and includes six reconfigurable thermo-optic phase shifters. We show that the device achieves high fidelity of operation throughout the full dynamical range. Then, we demonstrate experimentally the capability to achieve quantum-enhanced performances in multiphase estimation by using two-photon input states with respect to classical strategies, post-selected to the number of detected coincidences. Finally, we show that the same device can be employed to tune the input and output transformation to investigate the role of measurement operators in this scenario. The device reconfigurability can be exploited to implement general adaptive multiphase estimation protocols \cite{granade,piccoloLume,paesani}, thus providing a promising platform to develop appropriate methodologies for this task.

\section{Fabrication and characterization}

The integrated device, working at 785 nm, is composed as shown in Fig.~\ref{fig:device}a. The input state is prepared by a first unitary ($U^A$), where a reconfigurable thermo-optic phase shifter is employed to perform fine tuning of the implemented transformation. Then, the prepared state propagates through three internal waveguides with dynamical control of two independent phases between the three paths, ensured by four thermo-optic phase shifters. Finally, the output state undergoes a second unitary transformation $U^B$, implemented with the same layout of $U^A$, that is employed at the measurement stage. When the reconfigurable phases of $U^A$ and $U^B$ are set to $\pm \pi/2$, they act as balanced tritter, and the devices permit to engineer a reconfigurable 3-mode Mach-Zehnder interferometer (see Supplementary Material).

\subsection{Fabrication}
The photonic chip was fabricated by FLW, adopting a Yb-based cavity-dumped femtosecond laser oscillator operating at the output repetition rate of 1~MHz, and producing laser pulses of 300~fs duration and 1030~nm wavelength. The substrate employed was a commercial borosilicate glass (EagleXG, from Corning). The irradiation parameters used for the waveguides inscription are 250~nJ pulse energy and 30~mm/s substrate scan speed. The laser beam was focused 30 $\mu$m beneath the sample top surface by means of a microscope objective with 0.6 NA. The waveguide shallow depth was chosen to obtain an efficient control of the light phase by means of thermal shifters positioned on top of the circuit. The polarization of the writing beam was linear and set perpendicular with respect to the sample translation direction. With this fabrication configuration we obtained single-mode waveguides at the operating wavelength of 785~nm, with $1/e^2$ mode diameter of 7.2~$\mu$m~$\times$~8.4~$\mu$m and propagation losses $< 0.8$~dB/cm for the vertically polarized mode.

The thermal shifters that control the tritter operation and the interferometric phases were added to the photonic chip following the procedure presented in \cite{Flam15}. A thin and uniform gold layer (thickness of $\approx$~60~nm) was sputtered on top of the glass sample after the inscription of the waveguides. The gold layer was then patterned by FLW, in order to define the electrical circuit and the resistors above the waveguides, used as local heaters. As irradiation parameters, we used the second harmonic (at 515~nm) of the same laser employed for the waveguide writing, focused with a 0.6~NA objective on the glass surface, with a pulse energy of 100 nJ~and a scan speed of 2~mm/s. Each ablated line is scanned 8 times, in order to avoid parasitic shortcuts within the circuit. The resistors are fabricated with a width of 100~$\mu$m and a length in the waveguide direction ranging from 5~mm to 7~mm , which give resistance values in the range 60-100~$\Omega$. Standard electrical pins were directly glued on top of the circuit terminations, in order to facilitate the connectorization of our device with external power supplies.

\subsection{Characterization of the device}

As a first step after the fabrication process, we have characterized the integrated interferometer, in order to determine the relevant static parameters (beam-splitter transmittivies and internal phases when no voltage is applied) and the dynamical response of the device. In detail, a voltage $V_{R_i}$ applied on the resistor $R_i$ produces a power dissipation $P_{R_i}=V_{R_i}^{2}/R_{i}$; the temperature gradient in the chip induces a different phase shift along each optical path. Considering the combined action of multiple resistors, the resulting phase shifts are given by: 
\begin{equation}
\Delta\phi_{j}=\sum_{i=1}^6 \left(\alpha_{ji}P_{R_i}+\alpha_{ji}^{NL}P_{R_i}^{2} \right),
\end{equation}
where $\Delta\phi_{j}\ (j=1,2)$ are variations of the two independent phases of the three-arm interferometer (see inset in Fig. \ref{fig:device}), namely $\Delta \phi_1 = \varphi_1 - \varphi_{\mathrm{ref}}$ and $\Delta \phi_2 = \varphi_2 - \varphi_{\mathrm{ref}}$. Furthermore, $\alpha_{ij}$ and $\alpha_{ij}^{NL}$ are respectively the linear and non linear response coefficients associated to the dissipation $P_{R_i}$. The linear terms depend on all the geometric, thermal, and optical properties of the device \cite{Flam15}, while non-linear terms are associated to variations in the resistance value due to temperature increase.

The characterization procedure has been performed with single-photon inputs, generated by exploiting a $785$ nm photon-pair SPDC source, consisting in a type-II BBO nonlinear crystal pumped by a $392.5$ nm wavelength pump field (see Fig. \ref{fig:device}b-c). This allowed to measure the input-output probabilities $P(i \rightarrow j)$ from input $i$ to output $j$ . A detailed explanation of the full procedure and the corresponding results are reported in Supplementary Material. The high quality of operation of the device is confirmed by the average fidelity of the device, calculated using the characterized parameters, with respect to the set of achievable transformation in which both the tritters are considered to be ideal. Indeed, the fidelity $\langle F \rangle_{\Delta \phi_1, \Delta \phi_2}$, averaged over the interferometer phase differences $(\Delta \phi_1, \Delta \phi_2)$, reaches a value $\langle F \rangle_{\Delta \phi_1, \Delta \phi_2} = 0.963 \pm 0.015$. Here the fidelity is defined as $F = \vert \mathrm{Tr}[\tilde{U}(\Delta \phi_1, \Delta \phi_2) U^{\dag}(\Delta \phi_1, \Delta \phi_2)] \vert/m$, where $U(\Delta \phi_1, \Delta \phi_2)$ and $\tilde{U}(\Delta \phi_1, \Delta \phi_2)$ are respectively the ideal and reconstructed transformation for phases $(\Delta \phi_1, \Delta \phi_2)$. By exploiting the results of the characterization process, it is possible to control arbitrary phase differences between the interferometer arms by applying a suitable voltage on resistors $R_{i}$.

\section{Multiphase estimation on a chip}

After performing the characterization process, two-photon measurements have been performed as a function of phase differences $\Delta \phi_1$ and $\Delta \phi_2$, by setting transformations $U^{A}$ and $U^{B}$ as balanced tritters. Phases are tuned by varying voltages applied to resistors $R_1$ and $R_2$. The results are shown in Fig. \ref{3Dplot} and are compared with the theoretical predictions based upon the fit parameters obtained from the characterization process. Two-photon inputs are obtained by injecting both photons generated by the source into the integrated device (see Fig. \ref{fig:device}b-c). Two-photon coincidences are then recorded between the output detectors of the chip. The indistinguishability of the photon pairs injected into the chip was estimated from the visibility of a Hong-Ou-Mandel interference experiment, which gave the value $V = 0.95 \pm 0.01$.

\begin{figure}[ht!]
\centering
\includegraphics[width=0.48\textwidth]{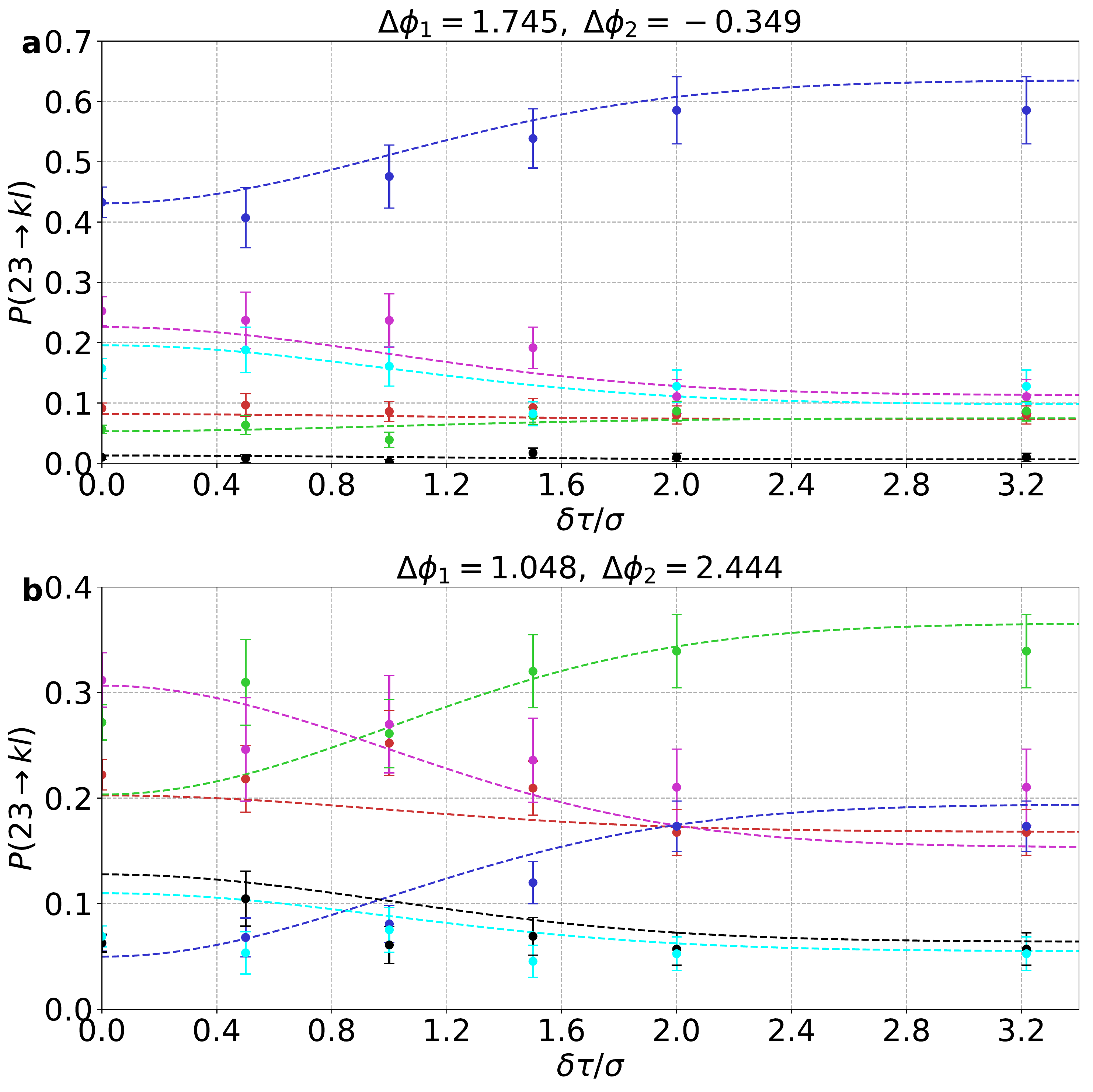}
\caption{Two-photon measurements $P(23 \rightarrow kl)$ for an input state with a single photon on modes (2,3) as a function of the relative time delay $\delta \tau$, normalized over the photon Hong-Ou-Mandel width $\sigma$. {\bf a}, Phase values set at $\Delta \phi_1 = 1.745$ and $\Delta \phi_2 = -0.349$. {\bf b}, Phase values set at $\Delta \phi_1 = 1.048$ and $\Delta \phi_2 = 2.444$. Points are experimental data, while dashed lines are predictions from the reconstructed parameters. [Red: output (1,2), Green: output (1,3), Blue: output (2,3), Black: output (1,1), Cyan: output (2,2), Purple: output (3,3)]. Photon indistinguishability is introduced in the predictions by mixing the probability with indistinguishable and distinguishable photons with a parameter $e^{-(\delta \tau/\sigma)^2}$. Error bars are standard deviations due to the Poissonian statistics of the measured two-photon coincidences.}
\label{transition}
\end{figure}

\begin{figure*}[ht!]
\centering
\includegraphics[width=1\textwidth]{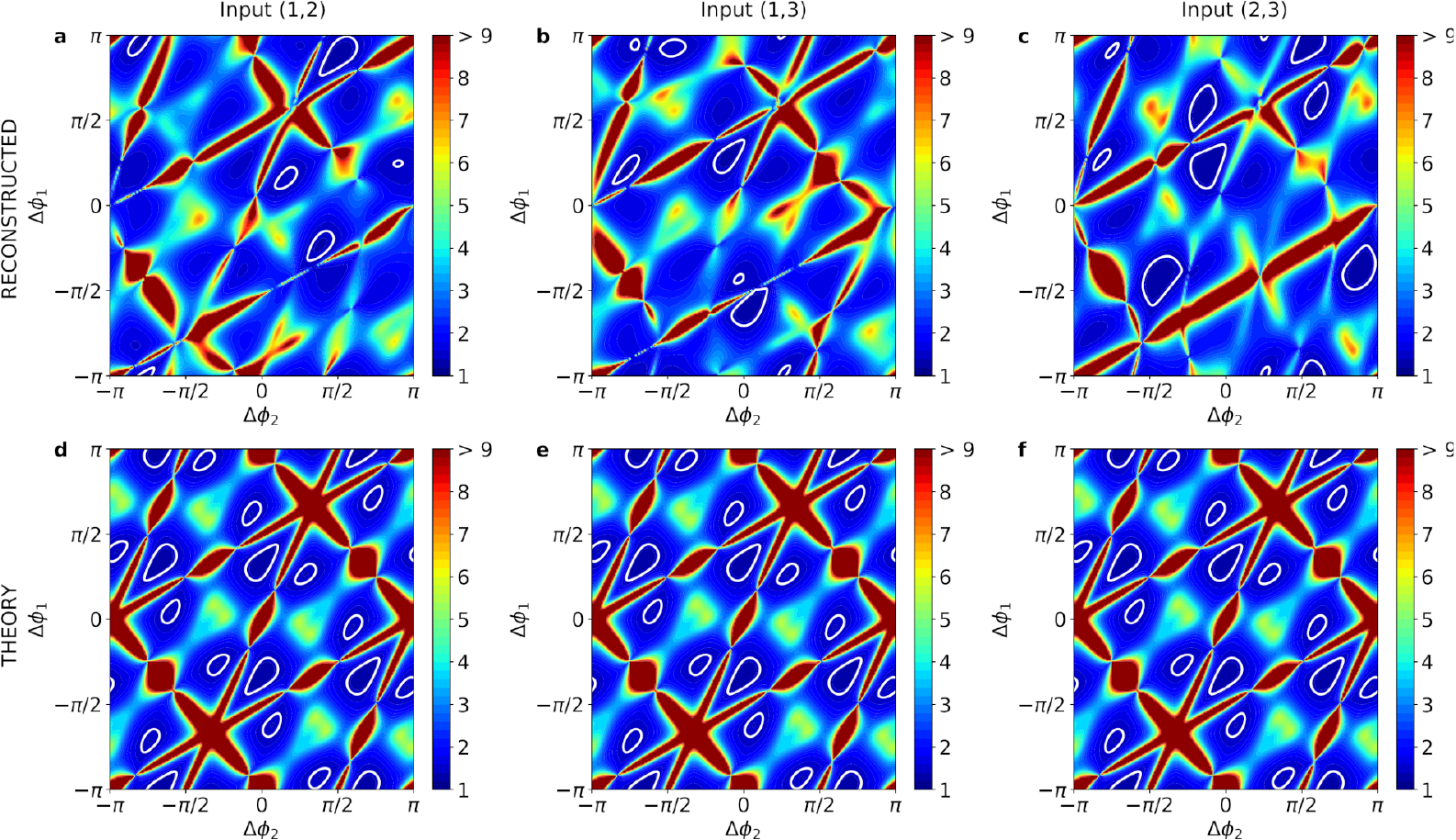}
\caption{Cramer-Rao bound $\mathrm{Tr}(\mathcal{I}^{-1})$ for multiphase estimation with two-photon input states. {\bf a-c}, CRB for the implemented device evaluated from the reconstructed parameters. {\bf a}, Input (1,2), {\bf b}, Input (1,3) and {\bf c}, Input (2,3). {\bf d-f}, CRB for the ideal three-mode interferometer. {\bf d}, Input (1,2), {\bf e}, Input (1,3) and {\bf f}, Input (2,3). In the ideal interferometer case, points where the Fisher information matrix is singular are not shown. Regions included within white closed curves highlight the presence of improved performances with respect to the QCRB with two distinguishable single-photon inputs.}
\label{fig:FIM}
\end{figure*}

As shown in Fig. \ref{3Dplot}, the measured two-photon output probabilities present a very good agreement with the theoretical model obtained by the reconstruction process. This demonstrates the capability to control the device transformation by simultaneously operating on multiple thermo-optic phase shifters (additional independent single-photon measurements are reported in Supplementary Material). The correct operation of the device is also confirmed by the capability of preserving quantum coherence during the evolution. In Fig.~\ref{transition} we show the coincidence detection measurements with two-photon inputs, as a function of the relative time delay $\delta \tau$. The latter is varied through adjustable delay lines, thus allowing to tune the degree of indistinguishability between the two input particles. The reported data show a clear signature of quantum interference when tuning the regime from indistinguishable to distinguishable particles.

\subsection{Experimental multiphase estimation} 

The present device can be directly employed to test and develop multiphase estimation protocols able to reach quantum-enhanced performances.  When dealing with multiphase estimation in a $(n+1)$-mode multiarm interferometer, the unknown parameters $\bm{\Phi}=(\Delta \phi_1, \Delta \phi_2,...,\Delta \phi_{n})$ are the $n$ independent phases relative to a reference arm. To perform their simultaneous estimation, an initial state $\rho_{0}$, is prepared by a unitary transformation $U^A$ and evolves through a transformation $U_{\bm{\Phi}}$ that encodes the information on the phases. Then, such information is extracted by a measurement $\hat{\Pi}_{x}$. Finally a suitable estimator $\bm{\hat{\Phi}}(\bm{x})=(\hat{\Delta \phi_1}(\bm{x}), \hat{\Delta \phi_2}(\bm{x}),...,\hat{\Delta \phi_{n}}(\bm{x}))$ provides an estimate of the phases by exploiting the $m$ measurement results $\bm{x}=(x_1,...,x_m)$. The phase sensitivity of an estimator, given a choice of the measurement operators, is quantified by its covariance matrix: $C(\bm{\Phi})_{ij}=\sum_{\bm{x}}[\bm{\hat{\Phi}}(\bm{x})-\bm{\Phi}(\bm{x})]_{i}\:[\bm{\hat{\Phi}}(\bm{x})-\bm{\Phi}(\bm{x})]_{j} \; P(\bm{x}|\bm{\Phi})$, with $i,j=1,...,n$. The precision in multiparameter estimation experiment can be evaluated as the trace of the covariance matrix, corresponding to the sum of the individual errors, in the form of the following chain of inequalities \cite{helm}: 
\begin{equation}
\label{ineq}
\sum_{i=1}^{n}\mathrm{Var}(\Delta \phi_i) \geq\;\frac{\mathrm{Tr}\left[ \mathcal{I}^{-1}(\bm{\Phi})\right]}{m}\geq\;\frac{\mathrm{Tr}\left[ \mathcal{H}^{-1}(\bm{\Phi})\right]}{m},
\end{equation}

where $\mathcal{I}(\bm{\Phi})$ is the Fisher information matrix, $\mathcal{H}(\bm{\Phi})$ is the quantum Fisher information matrix, and $m$ is the number of measurements. Optimal precision is achieved when the equality is saturated. In a more general form, such inequality should be written in matrix form as $C(\bm{\Phi})\ge \frac{\mathcal{I}^{-1}(\bm{\Phi})}{m} \ge \frac{\mathcal{H}^{-1}(\bm{\Phi})}{m}$, where the chain of inequality defines respectively the Cramer-Rao (CRB) and the quantum Cramer-Rao (QCRB) bounds. Indeed, as shown recently in \cite{Gessner}, the full covariance matrix has to be considered for a complete treatment of the sensitivity bounds. In particular, a one-by-one comparison between a desired target scenario (described by a Fisher information matrix $\hat{\mathcal{I}}$) and a given benchmark state (described by $\hat{\mathcal{H}}$) can be performed by calculating the number of positive eigenvalues of $\hat{\mathcal{I}}$ - $\hat{\mathcal{H}}$. This analysis provides the number of independent combination of the parameters where a sensitivity enhancement can be achieved by using the target state.

\begin{figure}[ht!]
\centering
\includegraphics[width=0.49\textwidth]{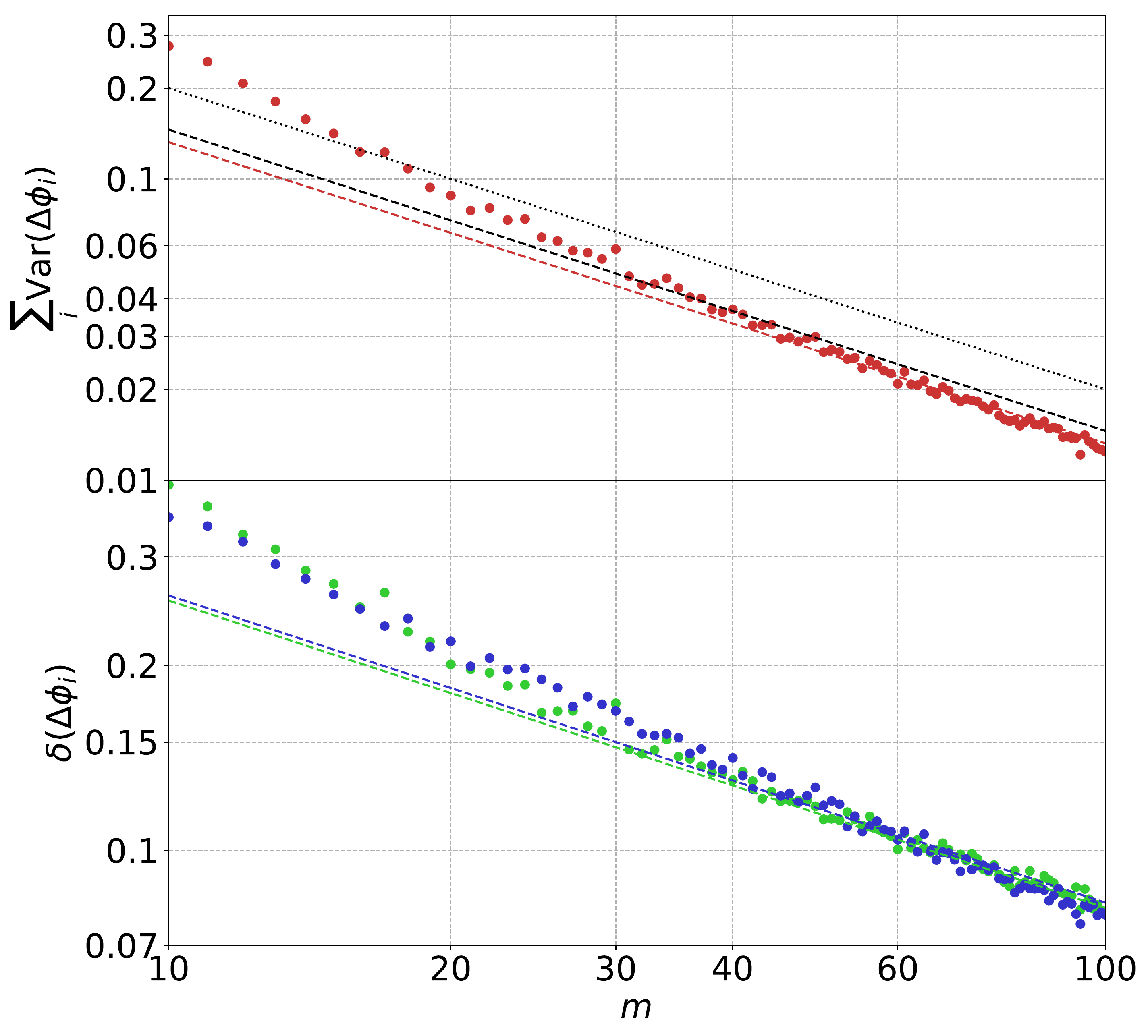}
\caption{Results of a maximum likelihood estimator for local phase estimation at $(\Delta \phi_1, \Delta \phi_2) = (-1.159, 2.810)$ with input (2,3). Points: experimental data, obtained by averaging over 100 random sequences of $m$ coincidence events ($2m$ photons) drawn from the measured $N_{\mathrm{ev}} = 1230$ two-photon events. Top plot: red dashed line corresponds to $\mathrm{Tr}(\mathcal{I}^{-1})$, black dashed line to the optimal sensitivity $\mathrm{Tr}(\mathcal{H}^{-1})$ with $2m$ distinguishable single-photon inputs, black dotted line to the optimal sensitivity when the phases are estimated separately with classical inputs. Bottom plot: green points (data) and line $(\mathcal{I}^{-1})^{1/2}_{11}$ correspond to $\delta (\Delta \phi_1)$, blue points (data) and line $(\mathcal{I}^{-1})^{1/2}_{22}$ correspond to $\delta (\Delta \phi_2)$.}
\label{Error}
\end{figure}

To verify the performance of the implemented device in this scenario, we have then evaluated the Fisher Information matrix $\mathcal{I}$ with two-photon inputs from the reconstructed parameters. Here, transformations $U^{A}$ and $U^{B}$ are set as balanced tritters. The results are shown in Fig. \ref{fig:FIM}, and compared with the corresponding calculations from an ideal three-mode balanced interferometer. We observe that, for all input states, regions can be identified that provide a value of $\mathrm{Tr}(\mathcal{I}^{-1})$ lower than the optimal bound achievable with two  distinguishable single-photon inputs, quantified by the corresponding matrix $\mathcal{H}$ \cite{Spagnolo13,Humphreys,ciamp}. While regions corresponding to quantum-enhanced performances are smaller than for ideal device, the minimum of $\mathrm{Tr}(\mathcal{I}^{-1})$ achieved by the implemented interferometer is close to the ideal value. Nevertheless, by exploiting adaptive protocols such performances can be extended to all pairs of phases if only a single region performs better than with classical resources.

We have then verified that enhanced estimation can be actually achieved by using an appropriate estimator. In Fig. \ref{Error} we show the results for a maximum likelihood (ML) approach in a local estimation framework for $(\Delta \phi_1, \Delta \phi_2) = (-1.159, 2.810)$ and input (2,3). The ML approach provides an estimate of the phases by maximizing the likelihood function $\mathcal{L}(\bm{\Phi}) = \prod_{k,l} P(23 \rightarrow kl)^{n_{kl}}$, where $n_{kl}$ is the number of measured events on output $(k,l)$. We observe that the overall error on both parameters, quantified by $\sum_i \mathrm{Var}(\Delta \phi_i)$, drops below the bound with the optimal separable inputs. More specifically, the achieved performance, exploiting $m$ two-photon events ($2m$ total photons) overcomes the scenario in which the phases are estimated simultaneously ($\mathcal{H}_{\mathrm{sim}}$) or separately ($\mathcal{H}_{\mathrm{sep}}$) with classical inputs having the same overall number of photons ($2m$) \cite{Humphreys}. The obtained enhancement with respect to a classical input is achieved in a post-selected scenario. Furthermore, the estimation of both parameters is achieved with comparable errors, thus leading to a symmetric estimation of the two phases. Finally, we can also compare the sensitivities relative to the full covariance matrices by using the approach of \cite{Gessner} discussed above. More specifically, we find that our system ($\mathcal{I}$) permits, for some pairs of phases, to obtain a sensitivity enhancement in both the two linearly independent combinations of the phases with respect to the scenario where the parameters are estimated separately ($\mathcal{H}_{\mathrm{sep}}$) by means of classical probes and also with respect to optimal simultaneous classical estimation ($\mathcal{H}_{\mathrm{sim}}$). Indeed, both the matrix differences  $\mathcal{I}-\mathcal{H}_{\mathrm{sep}}$ and   $\mathcal{I}-\mathcal{H}_{sim}$ have two positive eigenvalues.

The obtained enhancement can be extended to all pairs of phases by considering the application of adaptive estimation protocols \cite{piccoloLume,ciamp,paesani,granade}. This can be achieved with our device by exploiting the additional resistors $R_3$ and $R_4$ present in the circuit. The capability of performing adaptive protocols is particularly crucial in this multiparameter scenario, where the achievement of optimal \cite{matsu,hels,yuen} or symmetric \cite{ciamp} errors in all parameters are not always achievable.

\subsection{Tuning input and output transformations}

\begin{figure}[ht!]
\centering
\includegraphics[width=0.49\textwidth]{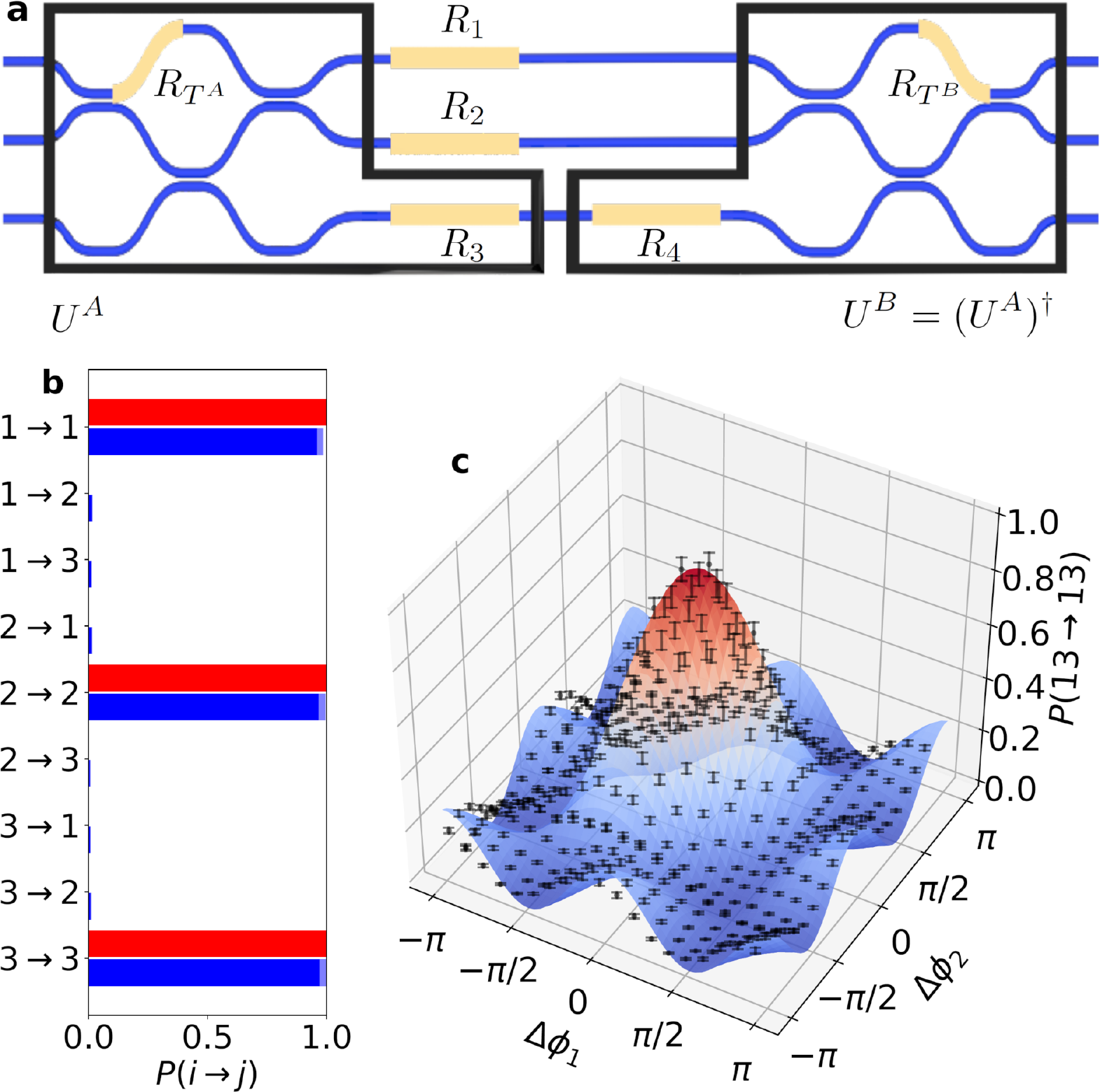}
\caption{{\bf a}, Conceptual layout employed to tune the input and output transformations $U^A$ and $U^B$  {\bf b}, Experimental single-photon probability measurements (blue bars) at $(\Delta \phi_1, \Delta \phi_2) = (0,0)$, compared with the identity corresponding to the ideal case (red bars). {\bf c}, Experimental two-photon probability measurements for input (1,3) and output (1,3) as a function of ($\Delta \phi_1, \Delta \phi_2$) by tuning voltages applied to resistors $R_1$ and $R_2$. {\bf b-c}, Transformations $U^{A}$ and $U^{B}$ are set to reach the condition $U^{B} U^{A} = I$ (up to a set of output phases) as described in the main text.}
\label{fig:UUcroce}
\end{figure}

The tunability of the device allows to implement different interferometers besides that composed by two cascaded tritters. This is obtained by varying the phases in resistors $R_{T^{A}}$ and $R_{T^{B}}$, and by exploiting the additional resistors $R_3$ and $R_4$ (see Supplementary Material for the characterization of resistors in $U^{A}$ and $U^{B}$). More specifically, let us consider the layout of Fig. \ref{fig:UUcroce}a. The additional phases on $R_3$ and $R_4$ are employed to configure the device such that $U^{B} U^{A} = I$ (up to a set of output phases). The implemented transformations $U^{A}$ and $U^{B}$, different from balanced tritters, are reported in Supplementary Material. This corresponds to tuning the device transformation as the identity $I$ for $(\Delta \phi_1, \Delta \phi_2) = (0,0)$. This scenario is particularly relevant in the multiparameter estimation context in order to saturate inequality (\ref{ineq}). Indeed, recent work \cite{pezze17}, providing the conditions for projective measurements to saturate such bound, has shown that such measurements include the projection over the initial state, thus requiring $U^{B} U^{A} = I$. The results are shown in Fig. \ref{fig:UUcroce}b-c. More specifically, we observe that the single-photon input-output probabilities $P(i \rightarrow j)$ closely resemble the identity matrix (see Fig. \ref{fig:UUcroce}b) at $(\Delta \phi_1, \Delta \phi_2) = (0,0)$, with a similarity $S = \frac{1}{3} \sum^3_{i=1}  P(i \rightarrow i) = 0.979 \pm 0.008$. Similar results are observed for two-photon inputs (see Fig. \ref{fig:UUcroce}c), thus showing the capability of tuning the input and output transformations by exploiting the additional phases embedded in the interferometer.

\subsection{Perspectives: improving sensitivity with multiphoton inputs}

\begin{figure}[ht!]
\centering
\includegraphics[width=0.35\textwidth]{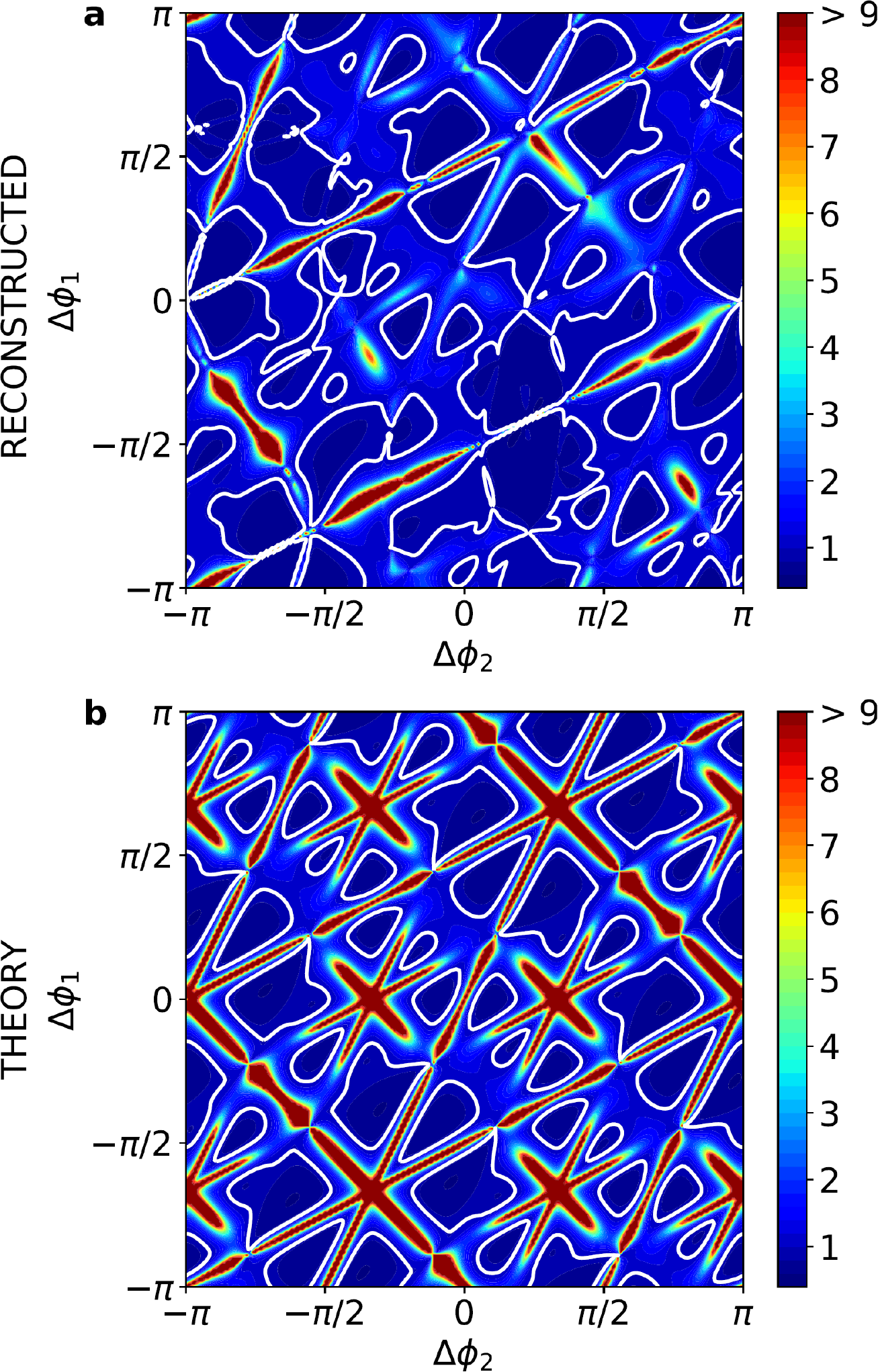}
\caption{Cramer-Rao bound $\mathrm{Tr}(\mathcal{I}^{-1})$ for multiphase estimation with a three-photon input state (1,2,3). {\bf a}, CRB for the implemented device evaluated from the reconstructed parameters, and {\bf b}, CRB for the ideal three-mode interferometer. In the ideal interferometer case, points where the Fisher information matrix is singular are not shown. Regions included within white closed curves highlight the presence of improved performances with respect to the QCRB with three optimal distinguishable single-photon inputs.}
\label{fig:three-photon}
\end{figure}

Sensitivity in multiphase estimation with the implemented device can be improved by changing the input state. For instance, let us consider a three-photon input where all modes are injected with a single photon. By evaluating the quantum Fisher information matrix $\mathcal{H}$ obtained after application of $U^A$ we obtain $\mathrm{Tr}(\mathcal{H}^{-1}) \simeq 0.527$, which is close to the value $0.5$ obtained for an ideal interferometer. The actual sensitivity after measuring the output state by applying transformation $U^B$ is quantified by the CRB $\mathrm{Tr}(\mathcal{I}^{-1})$, shown in Fig. \ref{fig:three-photon}. We observe that improved sensitivity can be achieved with the implemented device, leading to $\min \mathrm{Tr}(\mathcal{I}^{-1}) \simeq 0.584$, lower than the bound $\simeq 0.5+\sqrt{2}/3$ obtained by sending three distinguishable single photons prepared in the optimal state.

\section{Conclusions and discussion}

In order to fully disclose the potential of multiparameter estimation, several open problems are currently under investigation, both from the theoretical and the experimental side. In this context it is crucial to identify suitable platforms that can be employed to develop new methodologies and to benchmark their performances. 

In this article we have shown experimentally the capability of performing multiphase estimation in a reconfigurable integrated photonic chip realized via the femtosecond micromachining technology. Within such platform, the adoption of active thermo-optic phase shifters in a complex interferometric layout allows to investigate experimentally the simultaneous estimation of more than one Hamiltonian parameter. By properly tuning the input state, we have shown that such platform allows to reach quantum-enhanced performances with respect to what can be achieved with separable states, in a post-selected scenario to the number of detected coincidences. Furthermore, additional optical phase-shifters fabricated in the device increase the number of available control parameters. In this way we provide an experimental demonstration of a benchmark platform for the paradigmatic scenario of multiphase estimation in multimode interferometers.

Interesting perspectives can be envisaged starting from the presented results. On the one side, enlarging the dimensionality of the system will enable the investigation of a richer landscape \cite{ciamp}. On the other side, the capability of fabricating devices with additional controlled phases will allow to develop and test novel adaptive protocols \cite{piccoloLume,paesani,granade,ciamp}, or to tune the detection operator searching for the optimal measurement \cite{pezze17}. These ingredients can be combined in the same platform to develop a novel class of optimal protocols, allowing to efficiently extract information on an unknown set of parameters with minimal resource commitment.
Finally our platform is also suitable for the inclusion of other integrated elements allowing for all-in-chip processes: laser-written nonlinear waveguides, generating single photons \cite{Atzeni}, and microfluidic channels, enabling actual sensing experiments on fluid solutions \cite{crespiAPL}. This would allow all-in-chip multiphase estimation experiments, thus exploiting the potential of the platform.

\section*{Acknowledgments}
The authors thank A. S. Rab for support in the early stage of the experiment. This work is supported by the European Research Council (ERC) Advanced Grant CAPABLE (Composite integrated photonic platform by femtosecond laser micromachining, grant agreement no. 742745), and by the QuantERA ERA-NET Cofund in Quantum Technologies 2017 project HiPhoP (High dimensional quantum Photonic Platform, project ID 731473).

\end{document}


\title{Experimental multiphase estimation on a chip: supplementary material}

\author{Emanuele Polino}
\affiliation{Dipartimento di Fisica, Sapienza Universit\`{a} di Roma, Piazzale Aldo Moro 5, I-00185 Roma, Italy}

\author{Martina Riva}
\affiliation{Istituto di Fotonica e Nanotecnologie, Consiglio Nazionale delle Ricerche (IFN-CNR), Piazza Leonardo da Vinci, 32, I-20133 Milano, Italy}
\affiliation{Dipartimento di Fisica, Politecnico di Milano, Piazza Leonardo da Vinci, 32, I-20133 Milano, Italy}

\author{Mauro Valeri}
\affiliation{Dipartimento di Fisica, Sapienza Universit\`{a} di Roma, Piazzale Aldo Moro 5, I-00185 Roma, Italy}

\author{Raffaele Silvestri}
\affiliation{Dipartimento di Fisica, Sapienza Universit\`{a} di Roma, Piazzale Aldo Moro 5, I-00185 Roma, Italy}

\author{Giacomo Corrielli}
\affiliation{Istituto di Fotonica e Nanotecnologie, Consiglio Nazionale delle Ricerche (IFN-CNR), Piazza Leonardo da Vinci, 32, I-20133 Milano, Italy}
\affiliation{Dipartimento di Fisica, Politecnico di Milano, Piazza Leonardo da Vinci, 32, I-20133 Milano, Italy}

\author{Andrea Crespi}
\affiliation{Istituto di Fotonica e Nanotecnologie, Consiglio Nazionale delle Ricerche (IFN-CNR), Piazza Leonardo da Vinci, 32, I-20133 Milano, Italy}
\affiliation{Dipartimento di Fisica, Politecnico di Milano, Piazza Leonardo da Vinci, 32, I-20133 Milano, Italy}

\author{Nicol\`o Spagnolo}
\affiliation{Dipartimento di Fisica, Sapienza Universit\`{a} di Roma, Piazzale Aldo Moro 5, I-00185 Roma, Italy}

\author{Roberto Osellame}
\affiliation{Istituto di Fotonica e Nanotecnologie, Consiglio Nazionale delle Ricerche (IFN-CNR), Piazza Leonardo da Vinci, 32, I-20133 Milano, Italy}
\affiliation{Dipartimento di Fisica, Politecnico di Milano, Piazza Leonardo da Vinci, 32, I-20133 Milano, Italy}

\author{Fabio Sciarrino}
\affiliation{Dipartimento di Fisica, Sapienza Universit\`{a} di Roma, Piazzale Aldo Moro 5, I-00185 Roma, Italy}
\email{fabio.sciarrino@uniroma1.it}

\maketitle

This supplementary material contains a detailed theoretical analysis of the integrated device reported in the main text, and a full description of the experimental procedure for its characterization.

\section{Description of the integrated device} 

The implemented device is an integrated three-arm interferometer comprising two cascaded tritters,  built through the femtosecond laser-writing technique (FLW). A tritter is a three-port beam-splitter, and can be realized with different possible geometries. The FLW technique is suitable for a 3D realization of a tritter, that consists of a directional coupler made by three waveguides with a common coupling region \cite{Spagnolo13}. Nevertheless, in this article we realized the tritters embedded within the interferometer by employing a 2D geometry. Indeed, it is also possible to the decompose a tritter into cascaded two-mode beam splitters and a phase shifter \cite{Reck94}, as shown in Fig. 1 of the main text. This decomposition has the advantage of enabling the introduction of active reconfigurable elements to obtain fine control on the transformation.

The mathematical description of a tritter, in the lossless case, is given by a unitary matrix $U^{(3)}$ defined in the three-dimensional Hilbert space of the spatial modes. The matrix element of a symmetric tritter are $U^{(3)}_{i,i} = 3^{-1/2}$ and $U^{(3)}_{i,j} = 3^{-1/2} e^{\imath 2 \pi/3}$ for $i \neq j$. The evolution of the field operators is given by: $b_i^{\dagger}=\sum_{i,j}U^{(3)}_{ij}a_j^{\dagger} $, where $b_i^{\dagger}$ ($i=1,2,3$) are the output modes operators and the $a_j^{\dagger}$ ($j=1,2,3$) are the input modes operators.

\begin{figure*}[ht!]
\centering
\includegraphics[width=1\textwidth]{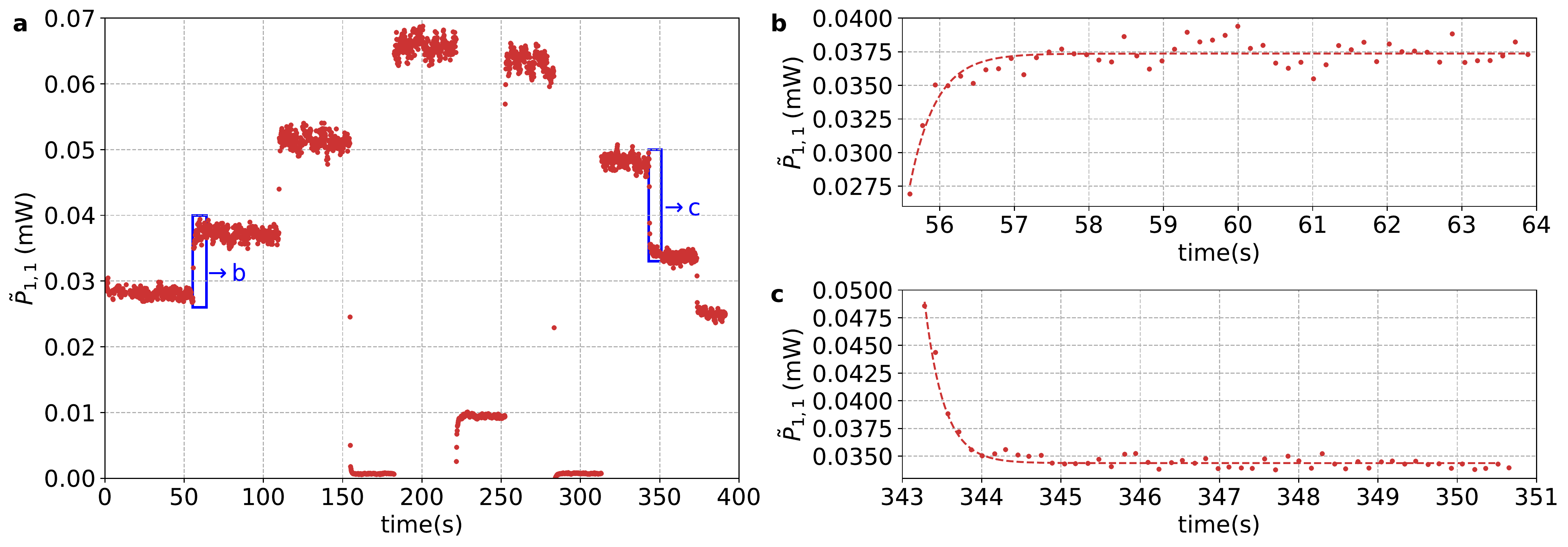}
\caption{{\bf a}, Characterization of thermo-optic shifters thermalization after tuning the applied voltage, by measuring the optical power $\tilde{P}_{1,1}$(mW) from output 1 after injecting classical light in input 1.  {\bf b}, zoom for interval $t \in [55.5;64]$s of transition corresponding to an increase in the optical power. {\bf c}, zoom for interval $t \in [343;352]$ s of transition corresponding to a decrease in the optical power. Points: experimental data. Dashed curves: best fit with an exponential decay model $f(t) = a + b e^{-(x-c)/\tau}$. In both insets, the measured time constant is $\tau \sim 0.3 s$.}
\label{fig:transient}
\end{figure*}

Let us now discuss the decomposition of the unitary matrix $U^A$ of the first tritter of the interferometer (Fig. 1 of main text). The first directional coupler mixes the first two modes of the interferometer. Assuming a lossless evolution, the reflectivity and transmission coefficients of the coupler $R^A_1$ and $T^A_1$ are related according to $R^A_1+T^A_1=1$. Hence, the directional coupler is described by the unitary matrix $U^{A}_{T_1}$:
\begin{equation}
U^A_{T_1} = \left( \begin{array}{cccc}
\sqrt{1-T^A_1} &i\sqrt{T^A_1} & 0 \\
i \sqrt{T^A_1} &\sqrt{1-T^A_1} & 0 \\
0 & 0 &1
\end{array} \right)\; .
\label{ut1}
\end{equation}
The second directional coupler mixes the second and the third modes and is described by the unitary matrix:
\begin{equation}
U^A_{T_2} = \left( \begin{array}{ccc}
1 & 0 &0\\
0&\sqrt{1-T^A_2} &i\sqrt{T^A_2}  \\
0& i \sqrt{T^A_2} &\sqrt{1-T^A_2}
\end{array} \right)\; ,
\label{ut2}
\end{equation}
where  $T^A_2$ is the transmission coefficient of the second directional coupler. To obtain the tritter transformation, an additional phase shifter $PS_{\varphi_{T^A}}$ that introduces a phase $\varphi_{T^A}$ between the first arm and the other two is required. Such transformation is described by the following matrix:
\begin{equation}
PS_{\varphi_{T^A}} = \left( \begin{array}{ccc}
e^{i \varphi_{T^A}}& 0 & 0\\
0 &1 & 0  \\
0&0 & 1
\end{array} \right)\; .
\label{fas}
\end{equation}
Finally the first two modes interfere again in a third directional coupler $U^A_{T_3}$, whose action is described by the same matrix of Eq. (\ref{ut1}) (with $T^A_3$ as the transmission coefficient). The first tritter is then described by a unitary matrix obtained as an appropriate product of the previously defined transformations:
\begin{equation}
U^{A}= U^{A}_{T_3} \cdot PS_{\varphi_{T^A}}  \cdot U^{A}_{T_2} \cdot U^{A}_{T_1} \;.
\label{tritt}
\end{equation}
The values of the transmission coefficients and the phase shift to obtain a symmetric tritter described by $U^{(3)}$ are: $T^{A}_1=T^{A}_3=1/2$, $T^{A}_2=2/3$ and $|\varphi_{T^A}|=\pi/2$. 

After the first transformation, the three phases embedded within the three internal arms of the interferometer (Fig.1 of main text) are described by the matrices:
\begin{equation}
\begin{aligned}
PS_1(\varphi_1)&=\begin{pmatrix}e^{i\varphi_1}&0&0\\0&1&0\\0&0&1\end{pmatrix}, \\
PS_2(\varphi_2)&=\begin{pmatrix}1&0&0\\0&e^{i\varphi_2}&0\\0&0&1\end{pmatrix}, \\
PS_3(\varphi_{\mathrm{ref}})&=\begin{pmatrix}1&0&0\\0&1&0\\0&0&e^{i\varphi_{\mathrm{ref}}}\end{pmatrix},
\end{aligned}
\end{equation}
where the latter term is chosen as the reference phase. 

The final transformation $U^{B}$ has the same form of $U^{A}$, Eq. (\ref{tritt}), with transmission coefficients  $T^{B}_1$, $T^{B}_2$, $T^{B}_3$ and internal phase $\varphi_{T^B}$.

The overall matrix $U^{\mathrm{int}}$ of the interferometer is: 
\begin{equation}
U^{\mathrm{int}}= U^{B} \cdot  PS_3(\varphi_{\mathrm{ref}}) \cdot PS_2(\varphi_2) \cdot PS_1(\varphi_1) \cdot  U^{A}\;.
\label{inter}
\end{equation} 
Therefore, the device is described by 11 parameters: the transmission coefficients of the tritters directional couplers $T^{A,B}_{1,2,3}$, phases $\varphi_{T^{A,B}}$ embedded in $U^{A,B}$, and the three internal phases $\varphi_1$, $\varphi_2$ and $\varphi_{\mathrm{ref}}$ of the interferometer. The output probabilities for single- or multi-photon states entering in the device can be thus calculated by using the evolution in Eq. (\ref{inter}).

The six transmission coefficients are fixed and depend on the coupling parameters between the involved waveguides. Conversely, all the phases can be tuned by thermo-optic phase shifters. Phases can be changed by applying appropriate voltages on the corresponding resistors according to Eq.(1) of main text. Six independent resistors are present in the structure. Resistors $R_{T^A}$ and $R_{T^B}$ are fabricated along the arms of the $U^{A,B}$ to directly change $\varphi_{T^A}$ and $\varphi_{T^B}$. Resistors  $R_{1-4}$ are distributed along the three arms between the two tritters as shown in Fig. 1 of main text. We observe that only two resistors (namely $R_1$ and $R_2$) are necessary to have the full control of the output probabilities. Indeed only two physically relevant phases can be identified, corresponding to the two phase differences with respect to the reference arm $\Delta \phi_1 = \varphi_{1} - \varphi_{\mathrm{ref}}$ and $\Delta \phi_2 = \varphi_{2} - \varphi_{\mathrm{ref}}$. As discussed in the main text, the other two resistors $R_{3}$ and $R_{4}$ can be employed to add further control on the device, enabling the implementation of adaptive protocols or the capability of tuning the measurement operators so as to have $U^{B} = (U^{A})^{\dag}$.

\section{Characterization of the integrated device} 

\subsection{Characterization of tunable phase shifts response time} 

\begin{figure*}[ht!]
\centering
\includegraphics[width=1\textwidth]{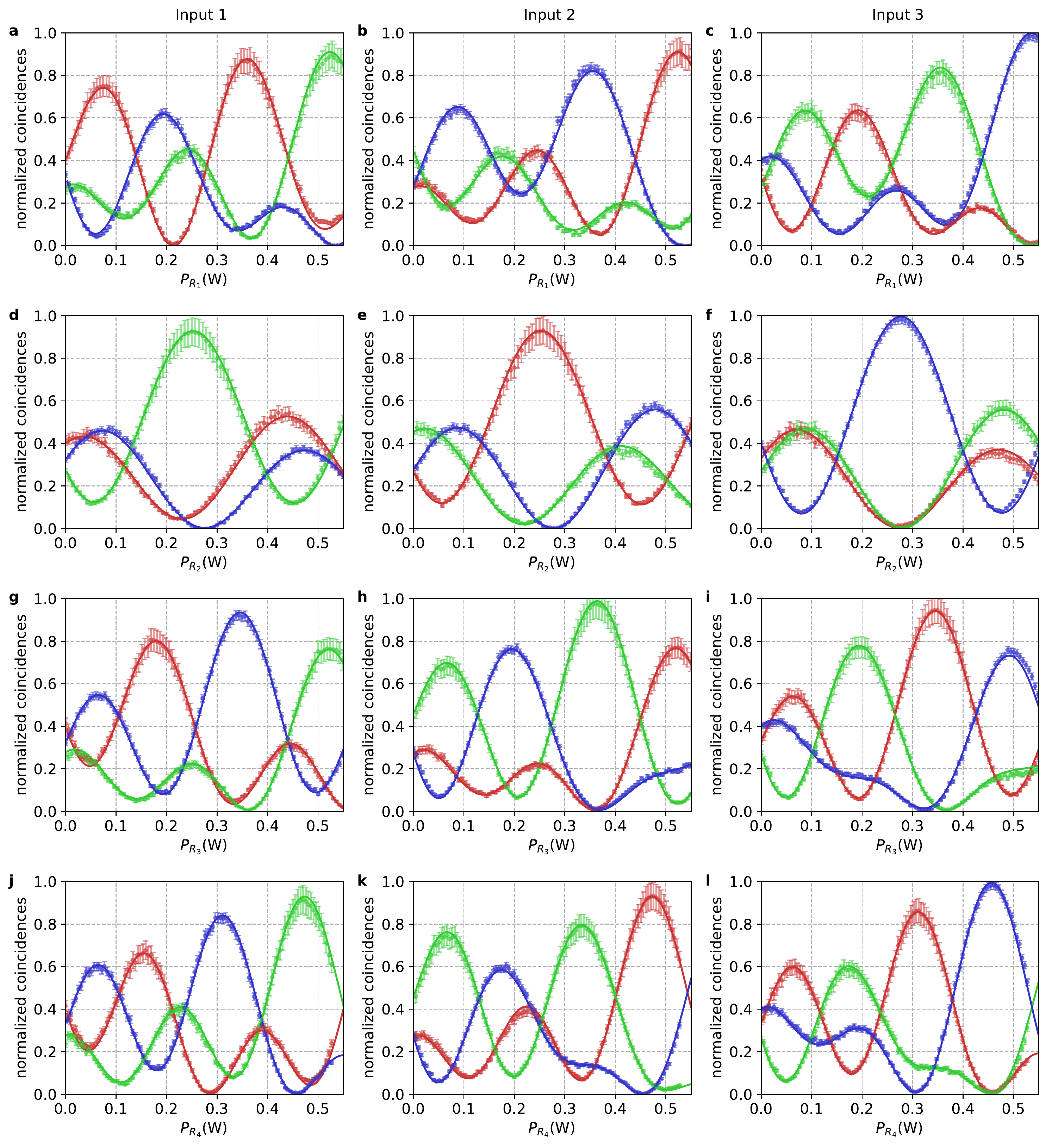}
\caption{Measured input-output probabilities (points) and relative fitted curves (solid lines) as a function of the dissipated power by resistor $R_1$ ({\bf a}-{\bf c}), $R_2$ ({\bf d}-{\bf f}), $R_3$ ({\bf g}-{\bf i}) and $R_4$ ({\bf j}-{\bf l}), where each resistor has been tuned separately. {\bf a,d,g,j}: Input 1. {\bf b,e,h,k}: Input 2. {\bf c,f,i,l}: Input 3. For each plot, output 1 corresponds to red points and lines, output 2 to green ones, and output 3 to blue ones. Those data are fitted to retrieve the values of the device parameters.}
\label{fig:fitR3R4}
\end{figure*}

As a first step towards characterization of the integrated device, it is necessary to measure the response time of the resistors after a modification of the applied voltage. This is required to evaluate thermalisation mechanisms of the thermo-optic resistors, and therefore to determine the waiting time after a voltage change to obtain a stable optical phase. This analysis is reported in Fig. \ref{fig:transient}. More specifically, we injected classical light in input 1 and measured the time evolution of the optical power emerging from output 1 by changing the applied voltage during the acquisition. The overall results are reported in Fig.\ref{fig:transient} a, while in Fig.\ref{fig:transient} b-c we focus on specific intervals. The transient regime can be described by an exponential decay model, leading to a time constant $\tau \sim 0.3$ s for our device. This constant includes all relevant transient mechanisms leading to a phase change in the device. To ensure a complete thermalisation of the chip during measurements, we choose a waiting time after voltage changes of $4$ s $\gg \tau$ before starting a new acquisition.

\subsection{Definition of the tritter phases} 

In the first stage of the characterization, according to the notation in Fig. 1 of the main text, we are interested in setting the phases $\varphi_{T^A}$ and $\varphi_{T^B}$ close to $\pm \pi/2$ to implement two ideal balanced tritters at $U^{A}$ and $U^{B}$. In this case, we exploit knowledge on the structure of $U^{\mathrm{int}}$ to search for appropriate input-output configurations for this characterization stage. Therefore we initially assume ideal coupling parameters for the device only to set the tritters phases. This is performed by following the three steps described below.

\textit{Step 1 (Interferometer phase setting). --}  By measuring $P(3 \rightarrow 3)$, the output probability is independent of both phases $\varphi_{T^A}$ and $\varphi_{T^B}$, since path leading to this input-output configuration do not propagate through the associated optical modes.  Therefore this condition can be used to set the internal phases to known values. The analytical expression of the output probability $P(3\rightarrow 3)$ for an ideal interferomter in this configuration is:
\begin{equation}
\begin{aligned}
P(3\to 3)&=\frac{1}{9}\{3-2 \cos(\varphi_1-\varphi_2)+2 \cos(\varphi_1-\varphi_{\mathrm{ref}})+\\
&-2 \cos(\varphi_2-\varphi_{\mathrm{ref}})\} \;.
\end{aligned}
\end{equation}
The internal phase difference can be set to $\varphi_{1}-\varphi_{2}=\varphi_{2}-\varphi_{\mathrm{ref}}=\pm\pi/3$ by minimizing this output probability through tuning of the voltages applied to $R_{1}$ and $R_{2}$. In our device, this is achieved for voltages values $V_{R_1}=2.05$ V and $V_{R_2} = 2.01$ V.

\textit{Step 2 ($U^{B}$ transformation). --}  The input-output probability $P(3 \rightarrow 1)$ [or equivalently $P(3 \rightarrow 2)$], is independent of the phase $\varphi_{T^A}$. Therefore, this configuration can be used to set the phase $\varphi_{T^B}$ to a known value by tuning the internal phases to the values found in step 1. The analytical expression of the output probability in this configuration is:
\begin{equation}
\begin{aligned}
P(3\to 1 &| \varphi_{1}-\varphi_{2}=\varphi_{2}-\varphi_{\mathrm{ref}}=\pm\pi/3)=\\
&= \frac{1}{2} (1 \mp \sin(\varphi_{T^B}))\;.	
\end{aligned}
\end{equation}
Phase $\varphi_{T^B}$ is then set to $\pm \pi/2$  by minimizing such output probability through tuning of the voltage applied to $R_{T^{B}}$. This is achieved for a voltage value $V_{R_{T^B}}=5.94$ V.

\textit{Step 3 ($U^{A}$ transformation). --}  The final step is performed to determine the transformation $U^{A}$. This is done by using the internal phases and the value of $\varphi_{T^B}$ found in the previous steps 1-2. More specifically, by measuring $P(1 \rightarrow j)$ [or equivalently $P(2 \rightarrow j)$] for any $j$, this path now depends on phase $\varphi_{T^A}$. For instance, the analytical expression of $P(1\to 1)$ in this configuration is:
\begin{equation}
\begin{aligned}
P(1\to 1 &|\varphi_{1}-\varphi_{2}=\varphi_{2}-\varphi_{\mathrm{ref}}=\pm\pi/3 , \varphi_{T^B}=\pm \pi/2)= \\
&=\frac{1}{2} (1 \pm \sin(\varphi_{T^A}))\;.
\end{aligned}
\end{equation}
Phase $\varphi_{T^A}$ is then set to $\pm \pi/2$  by minimizing such output probability through tuning of the voltage applied to $R_{T^{A}}$. This is achieved for a voltage value $V_{R_{T^A}}=2.90$ V.

In conclusion, the transformations $U^{A}$ and $U^{B}$ correspond to balanced tritters for voltages $V_{R_{T^A}}=2.90$ V and $V_{R_{T^B}}=5.94$ V. Hence, the multiarm integrated interferometer chip implements a balanced reconfigurable three-mode Mach-Zehnder interferometer.

\begin{table}[b!]
\centering
\def\arraystretch{1.5}
\begin{tabular}{|c|c|}
\hline
Parameter & Estimated value \\ 
\hline
$\overline{\Delta \phi}_{10}$ (rad) & $-0.355 \pm 0.005$ \\
\hline
$\overline{\Delta \phi}_{20}$ (rad) & $-1.441 \pm 0.004$ \\
\hline
$\overline{\varphi}_{T^A}$ (rad) & $1.893 \pm 0.002$ \\
\hline
$\overline{\varphi}_{T^B}$ (rad) & $1.866 \pm 0.002$ \\
\hline
$\overline{T}_{1}^A$ & $0.414 \pm 0.001$ \\
\hline
$\overline{T}_{1}^B$ & $0.415 \pm 0.001$ \\
\hline
$\overline{T}_{2}^A$ & $0.617 \pm 0.001$ \\
\hline
$\overline{T}_{2}^B$ & $0.625 \pm 0.001$ \\
\hline
$\overline{T}_{3}^A$ & $0.411 \pm 0.001$ \\
\hline
$\overline{T}_{3}^B$ & $0.438 \pm 0.001$ \\
\hline
$\overline{\alpha}_{11}$ (rad W$^{-1}$) & $24.35 \pm 0.06$ \\
\hline
$\overline{\alpha}_{11}^{NL}$ (rad W$^{-2}$) & $-0.34 \pm 0.12$ \\
\hline
$\overline{\alpha}_{21}$ (rad W$^{-1}$) & $8.85 \pm 0.05$ \\
\hline
$\overline{\alpha}_{21}^{NL}$ (rad W$^{-2}$) & $-0.66 \pm 0.11$ \\
\hline
$\overline{\alpha}_{12}$ (rad W$^{-1}$) & $0.72 \pm 0.05$ \\
\hline
$\overline{\alpha}_{12}^{NL}$ (rad W$^{-2}$) & $-0.11 \pm 0.08$ \\
\hline
$\overline{\alpha}_{22}$ (rad W$^{-1}$) & $16.54 \pm 0.03$ \\
\hline
$\overline{\alpha}_{22}^{NL}$ (rad W$^{-2}$) & $-0.55 \pm 0.06$ \\
\hline
$\overline{\alpha}_{13}$ (rad W$^{-1}$) & $-23.54 \pm 0.05$ \\
\hline
$\overline{\alpha}_{13}^{NL}$ (rad W$^{-2}$) & $-0.16 \pm 0.10$ \\
\hline
$\overline{\alpha}_{23}$ (rad W$^{-1}$) & $-17.45 \pm 0.04$ \\
\hline
$\overline{\alpha}_{23}^{NL}$ (rad W$^{-2}$) & $-0.66 \pm 0.08$ \\
\hline
$\overline{\alpha}_{14}$ (rad W$^{-1}$) & $-26.65 \pm 0.06$ \\
\hline
$\overline{\alpha}_{14}^{NL}$ (rad W$^{-2}$) & $0.03 \pm 0.14$ \\
\hline
$\overline{\alpha}_{24}$ (rad W$^{-1}$) & $-17.20 \pm 0.05$ \\
\hline
$\overline{\alpha}_{24}^{NL}$ (rad W$^{-2}$) & $-1.21 \pm 0.11$ \\
\hline
\end{tabular}
\caption{Best fit values and corresponding errors of the relevant chip parameters obtained from the characterization procedure.}
\label{table:fit}
\end{table}

\subsection{Characterization of the thermal response coefficients of the internal phases} 

The response of the relevant phases ($\Delta\phi_j$, $j=1,2$) in the interferometer is described by Eq.(1) of the main text. Indeed, $\Delta\phi_j$ depends on thermal response coefficients $\alpha_{ji}$ and $\alpha_{ji}^{NL}$ of the internal modes, where $(j,i)$ indicates the action of a voltage $V_{R_i}$ applied on the resistor $R_i$. After setting the transformations $U^{A,B}$ to balanced tritters, it is necessary to determine the complete set of relevant chip parameters.  More specifically, it is necessary to reconstruct the thermal response coefficients of the internal phases and the relevant static parameters of the device (internal phases at zero applied voltage and transmittivities of the directional couplers). In order to estimate these coefficients, single photon measurements (see Fig. \ref{fig:fitR3R4}) are performed as follows. For each input $i$ and output $j$, probability $P(i \rightarrow j)$ is recorded by tuning the applied voltage to each resistor separately. In such a way, the action of each resistor is characterized independently. This procedure corresponds to an overall amount of 36 independent input-output probability measurements.

\begin{figure*}[ht!]
\centering
\includegraphics[width=1\textwidth]{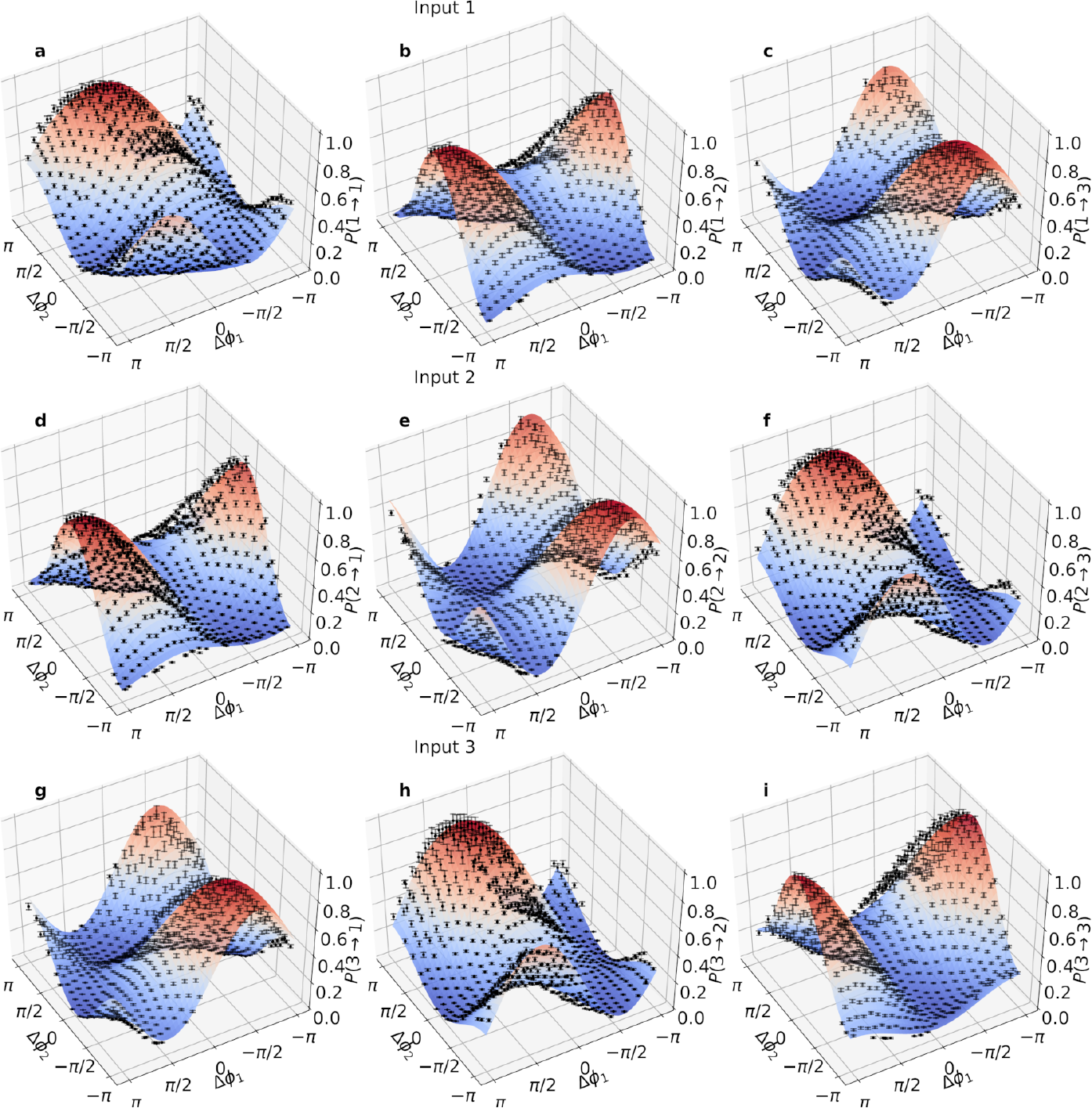}
\caption{Measured single-photon input-output probabilities $P(i \rightarrow j)$ as a function of phase differences $\Delta \phi_{1}$ and $\delta \phi_{2}$, tuned by simultaneously varying the dissipated power in resistors $R_{1}$ and $R_{2}$. Points: experimental data. Surfaces: curves obtained from the characterized parameters and from the model of Eq. (\ref{inter}). {\bf a-c}, Input 1, {\bf d-f}, input 2 and {\bf g-i}, input 3. For each input, the three plots correspond to the three different output modes. The good agreement between model and experimental data is quantified by the average $R^{2}$ value over all output combinations $\langle R^{2} \rangle = 0.965$.}
\label{fig:singlephoton}
\end{figure*}

\begin{figure*}[ht!]
\centering
\includegraphics[width=1\textwidth]{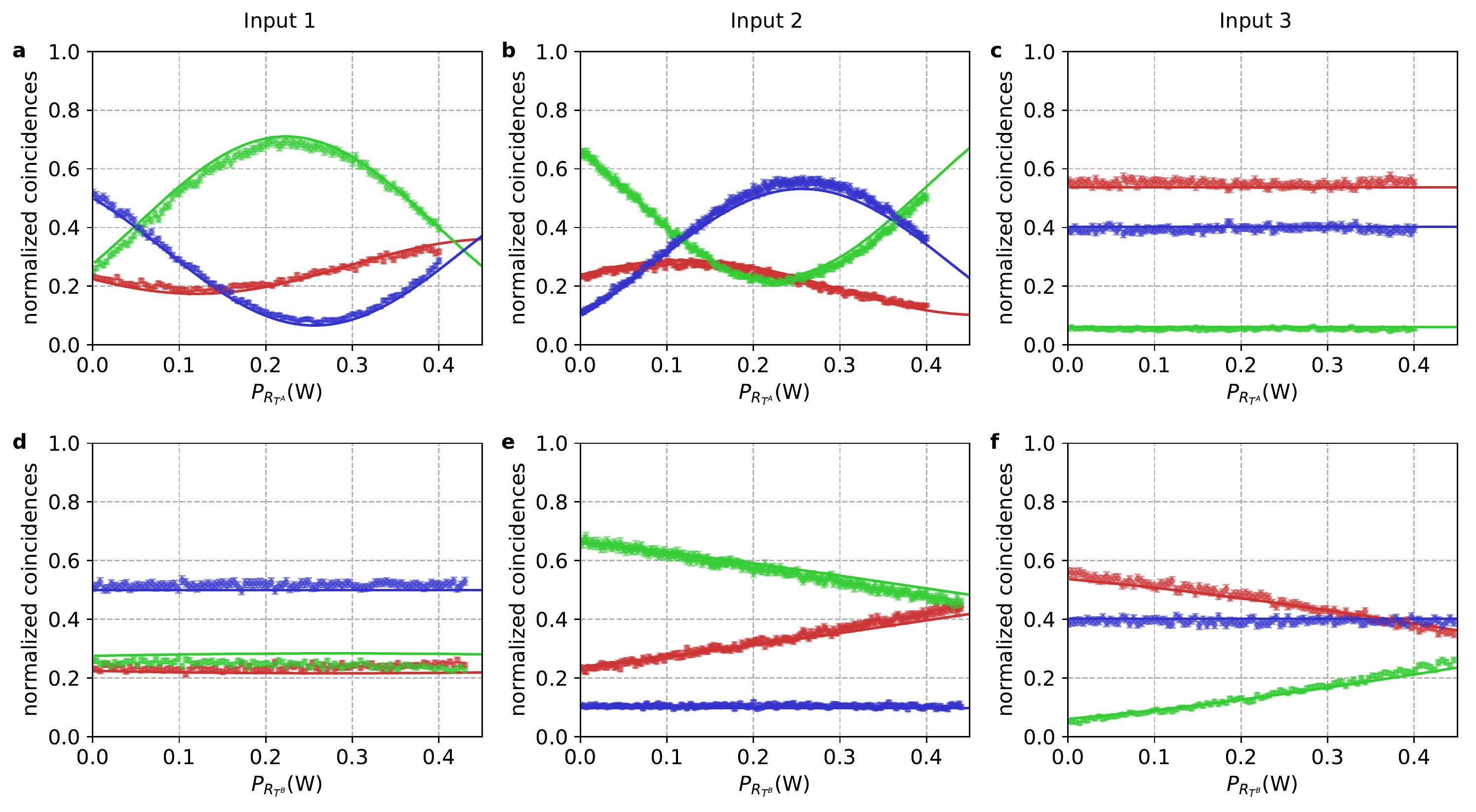}
\caption{Measured input-output probabilities (points) and relative fitted curves (solid lines) as a function of the dissipated power by resistor $R_{T^{A}}$ ({\bf a}-{\bf c}) and $R_{T^{B}}$ ({\bf d}-{\bf f}), where each resistor has been tuned separately. All other voltages are set to 0. {\bf a,d}: Input 1. {\bf b,e}: Input 2. {\bf c,f}: Input 3. For each plot, output 1 corresponds to red points and lines, output 2 to green ones, and output 3 to blue ones. Those data are fitted to retrieve the values of the relevant parameters for resistors $R_{T^{A}}$ and $R_{T^{B}}$ necessary to tune $U^{A}$ and $U^{B}$.}
\label{fig:fitRTARTB}
\end{figure*}

All reconstructed curves are then simultaneously fitted with the theoretical model $U^{\mathrm{int}}$ of Eq. (\ref{inter}). This fit estimate all the relevant parameters described above, and the two initial static relative phases $\overline{\Delta \phi}_{10}$ and $\overline{\Delta \phi}_{20}$ of the internal interferometer arms when no voltage is applied, for an overall amount of 26 parameters. During this procedure, it is necessary to give reasonable starting values for all these parameters to enable the correct convergence of the numerical minimization. Such starting values are obtained as follows. According to the fabrication process (see Fig. 1 of the main text) and to the first stage of the characterization, true values of the tritter parameters can be searched around their expected values:
\begin{eqnarray}
\overline{\varphi}_{T^{A,B}}&=&\pi/2+\Delta\varphi_{T^{A,B}},\\
\overline{T}^{A,B}_{1,3}&=&1/2+\Delta T^{A,B}_{1,3},\\
\overline{T}^{A,B}_{2}&=&2/3+\Delta T^{A,B}_{2}.
\end{eqnarray}
Conversely, as starting points for the thermal response coefficients we adopt the fundamentals harmonics obtained by performing Fourier analysis on the measured data.  Hence, the output probabilities are expressed as a linear combinations of sines and cosines, where the relative arguments depend on thermal response coefficients. Note that the number of harmonics generated from each measurement is bigger than the real one. This is observed since each curve is measured on a finite range of dissipated power. The analysis has been then restricted by keeping those harmonics shared by all curves. Moreover, to assign each harmonic value to the appropriate coefficient, we considered that each resistor has a greater influence on the nearest modes. Then, the obtained results are employed as a starting point for the fitting procedure.

The value of $\chi^2$ after the characterization process is evaluated as the squared difference between experimentally measured probabilities and the predictions obtained from the expected curve with best fit parameters, normalized to the errors on the data. We obtained $\chi^{2} = 3795$ with $\nu = 3258$ experimental points. All resulting parameter values are reported in Tab.~\ref{table:fit} with the associated errors. We observe that phases $\varphi_{T^A}$ and $\varphi_{T^B}$ are set to values $\overline{\varphi}_{T^A} = 1.893 \pm 0.002$ and $\overline{\varphi}_{T^B} = 1.866 \pm 0.002$ due to the non-ideality of the coupling parameters of the tritter directional couplers.  However, this difference with respect to ideal $\pi/2$ does not correspond to additional inaccuracy in the characterization process, since all actual device parameters are recovered by the fitting procedure.

\subsection{Verification of the characterization procedure by simultaneously tuning the interferometer phases $\Delta \phi_{1}$ and $\Delta \phi_{2}$}

After performing the characterization procedure, we have then verified that the reconstructed parameters are able to correctly predict the behaviour of the device and the value of the applied phases $\Delta \phi_{1}$ and $\Delta \phi_{2}$. To this end, we have collected an independent set of measurements with single-photon inputs, by varying both phases $\Delta \phi_{1}$ and $\Delta \phi_{2}$ simultaneously and by setting the transformation $U^{A}$ and $U^{B}$ as balanced tritters. The obtained results are shown in Fig. \ref{fig:singlephoton}, while two-photon measurements are reported in Fig. 2 of the main text. We observe that the curves calculated from the reconstructed parameters provide a good description of the measured experimental data obtained by simultaneously tuning the dissipated power in both resistors $R_{1}$ and $R_{2}$.

\subsection{Characterization of the thermal response coefficients of phases in $U^{A}$ and $U^{B}$} 

During the initial characterization, the phases on $R_{T^{A}}$ and $R_{T^{B}}$ ($\varphi_{T^A}$ and $\varphi_{T^B}$) are set to values near $\pm \pi/2$ to realize balanced tritter as $U^A$ and $U^B$. 
However, it is necessary to fully characterize the thermal response of those phases to exploit the full potential of the device. Therefore, we need to measure linear and non linear thermal response coefficients of the two phases $(\alpha_{T^A}, \alpha_{T^B}, \alpha_{T^A}^{NL}, \alpha_{T^B}^{NL})$. Single photon measurements (see Fig. \ref{fig:fitRTARTB}) have been performed as in the previous  characterization to estimate those coefficients, namely by separately tuning the applied voltage on resistors $R_{T^{A}}$ and $R_{T^{B}}$. This procedure corresponds to an overall amount of 18 independent input-output probability measurements. The set of unknown parameters to be determined also includes the two initial static phases $\overline{\varphi}_{0_{T^A}}$ and $\overline{\varphi}_{0_{T^B}}$ when no voltage is applied on the corresponding resistors. Finally, The measured data are processed to obtain a best fit of the 6 parameters. The obtained value of the $\chi^2$ is 6591 with 2418 experimental points. All resulted fitting values are reported in the Table~\ref{table:tritterfit}. Characterization of the dynamical response of resistors $R_{T^{A}}$ and $R_{T^{B}}$ can be exploited to perform a fine tuning of the phases $\varphi_{T^A}$ and $\varphi_{T^B}$. Furthermore, this characterization procedure has been then exploited to implement $U^{A}$ and $U^{B}$ transformations different from ideal balanced tritters (see Fig. 6 of the main text). 

\begin{table}[ht!]
\centering
\def\arraystretch{1.5}
\begin{tabular}{|c|c|}
\hline
Parameter & Estimated value \\
\hline
$\overline{\varphi}_{0_{T^A}}$ (rad) & $1.137 \pm 0.002$ \\
\hline
$\overline{\varphi}_{0_{T^B}}$ (rad) & $0.914 \pm 0.002$ \\
\hline
$\overline{\alpha}_{T^A}$ (rad W$^{-1}$) & $9.06 \pm 0.04$ \\
\hline
$\overline{\alpha}_{T^A}^{NL}$ (rad W$^{-2}$) & $-0.35 \pm 0.12$ \\
\hline
$\overline{\alpha}_{T^B}$ (rad W$^{-1}$) & $1.83 \pm 0.03$ \\
\hline
$\overline{\alpha}_{T^B}^{NL}$ (rad W$^{-2}$) & $0.75 \pm 0.09$ \\
\hline
\end{tabular}
\caption{Best fit values and corresponding errors of parameters which describe the thermal response of the phases in $U^{A}$ and $U^{B}$.}
\label{table:tritterfit}
\end{table}

\subsection{Implemented transformations $U^A$ and $U^{B}$}

Here we report the implemented transformations $U^{A}$ and $U^{B}$ for the two-photon measurements reported in the main text. The integrated chip structure allows to tune its parameters $\varphi_{T^{A}}$ and $\varphi_{T^{B}}$ in order to set the input and output transformation as balanced tritters, as described by the characterization procedure discussed above. By setting those parameters to the values discussed in Table \ref{table:fit}, and by considering the characterized static values of $T_{1,2,3}^{A,B}$ reported in the same table, the transformations $U^{A}$ and $U^{B}$ are set respectively to:
\begin{equation}
U^{A} = \begin{pmatrix} 
-0.441 + 0.557 \imath & -0.468 + 0.148 \imath & -0.504 \\ 
-0.466 + 0.150 \imath & 0.494 - 0.391 \imath & 0.602 \imath \\
-0.505 & 0.601 \imath & 0.619
\end{pmatrix},
\end{equation}
and
\begin{equation}
U^{B} = \begin{pmatrix}
-0.428 + 0.549 \imath & -0.462 + 0.170 \imath & -0.523 \\
-0.484 + 0.149 \imath & 0.475 - 0.408 \imath & 0.592 \imath \\
-0.509 & 0.605 \imath & 0.613
\end{pmatrix}
\end{equation}
These unitaries present a high value of fidelity $F^{A,B} = \vert \mathrm{Tr}[\tilde{U}^{A,B}(U^{(3)})^{\dag}] \vert/m$, where $m=3$, with respect to a balanced tritter $U^{(3)}$, namely $F^{A} = 0.9830 \pm 0.0002$ and $F^{B} = 0.9863 \pm 0.0001$.

As discussed in the main text, the additional resistors $R_3$ and $R_4$ can be employed to tune the input and output transformations $U^{A}$ and $U^{B}$ to reach the condition $U^{B} U^{A} = I$ (up to a set of output phases). The measurements reported in Sec. 3B have been collected by setting the transformations $U^{A}$ and $U^{B}$ respectively as:
\begin{equation}
U^{A} = \begin{pmatrix}
-0.368 - 0.562 \imath & 0.476 + 0.220 \imath & -0.523 + 0.003 \imath \\
-0.499 + 0.202 \imath & 0.431 + 0.417 \imath & 0.002 + 0.592 \imath \\
-0.509 & 0.605 \imath & 0.613
\end{pmatrix},
\end{equation}
and
\begin{equation}
U^{B} = \begin{pmatrix}
-0.392 + 0.564 \imath & 0.452 - 0.266 \imath & -0.504 \\
-0.487 + 0.191 \imath & -0.390 + 0.460 \imath & 0.605 \imath \\
-0.505 + 0.0147 \imath & -0.080 - 0.596 \imath & 0.619
\end{pmatrix}.
\end{equation}